\documentclass[twocolumn]{aa}

\usepackage{mathptmx}
\usepackage[flushleft]{threeparttable}
\usepackage{longtable}
\usepackage{color}
\usepackage{graphicx}
\usepackage{amsmath}
\usepackage{amssymb}
\usepackage[T1]{fontenc}
\usepackage{mathptmx}
\usepackage{ae,aecompl} 
\usepackage[varg]{txfonts}

\begin{document}
\title{Toy model for the acceleration of blazar jets}
\author{I. Liodakis\inst{\ref{inst1}}\thanks{ilioda@stanford.edu}
}

\institute{KIPAC, Stanford University, 452 Lomita Mall, Stanford, CA 94305, USA\label{inst1}}

\abstract{Understanding the acceleration mechanism of astrophysical jets has been a cumbersome endeavor from both the theoretical and observational perspective. Although several breakthroughs have been achieved in recent years, on all sides, we are still missing a comprehensive model for the acceleration of astrophysical jets}{ In this work we attempt to construct a simple toy model that can account for several observational and theoretical results and allow us to probe different aspects of blazar jets usually inaccessible to observations.} { We used the toy model and Lorentz factor estimates from the literature to constrain the black hole spin and external pressure gradient distributions of blazars.}{ Our results show that (1) the model can reproduce the velocity, spin and external pressure gradient of the jet in M87 inferred independently by observations; (2) blazars host highly spinning black holes with 99\% of BL Lac objects and 80\% of flat spectrum radio quasars having spins $a>0.6$; (3) the dichotomy between BL Lac objects and Flat Spectrum Radio Quasars could be attributed to their respective accretion rates. Using the results of the proposed model, we estimated the spin and external pressure gradient for 75 blazars.}{}


\titlerunning{Acceleration of astrophysical jets}
\authorrunning{I. Liodakis}
\maketitle

\section{Introduction}\label{intro}

Black holes (BH) of all masses are capable of producing collimated relativistic plasma outflows called jets. These jets are most likely produced via the Blandford-Znajek mechanism (BZ mechanism, \citealp{Blandford1977}) where the energy powering the jet is extracted from the spin of the BH. Although the first jet was discovered a century ago in M87 \citep{Curtis1918}, the structure and acceleration mechanism of astrophysical jets remains an important unanswered question and field of active research to this day. In recent years great progress has been made in both theoretical and observational perspectives. Progress in the former is due to the increasing ability of modern computers to handle complex and computationally demanding simulations, while in the latter due to new facilities pushing the boundaries of energy and angular resolution. However, although progress has been made in different individual fields, we are still missing a unifying scheme for the structure and acceleration of BH-powered jets. Frequently used assumptions for the structure of the jets involve cylindrical, conical, and parabolic geometries, while the velocity of the jet ($u_j$, usually expressed in terms of the Lorentz factor $\Gamma=(1-(u_j/c)^2)^{-1/2}$) is often assumed to be constant throughout the jet. However, variability timescales from different regions of the jet would imply, in at least some sources, different beaming properties (e.g., \citealp{Ghisellini2005}) rendering the constant Lorentz factor scenario unlikely. Acceleration is therefore a necessary ingredient in the jet paradigm.

From the theoretical perspective thermal driving has been shown to be inadequate to explain the high $\Gamma$
 seen in jets suggesting that they have to initially be magnetically dominated \citep{Vlahakis2004,Vlahakis2015-book}. For magnetically dominated jets the external pressure from the surrounding medium has an important contribution to the acceleration process  \citep{Vlahakis2015-book}. This has also been demonstrated in analytical and numerical work by \cite{Komissarov2007,Komissarov2009} and  \cite{Lyubarsky2009,Lyubarsky2010}. In \cite{Lyubarsky2009,Lyubarsky2010} it is shown that the external pressure could be responsible for the collimation of Poynting dominated jets and that the collimation and acceleration can take place over large distances. The jets are efficiently accelerated in the ``equilibrium'' regime while the Poynting dominated jet is slowly converted to a matter-dominated jet. Although the jet will only become fully matter dominated at much larger distances, the acceleration is likely to stop when the magnetization parameter is $\sigma\leq 1$ \citep{Vlahakis2003,Vlahakis2004-II,Lyubarsky2009}. In the equilibrium regime, the jet will expand with decreasing external pressure until the pressure becomes constant. Then the jet will transition to a cylindrical geometry. Similar results have been obtained in \cite{Komissarov2007,Komissarov2009} where the magnetically dominated jet is confined by external pressure with a power-law profile ($p\propto z^{-s}$). The jet has a parabolic shape as long as the power-law exponent is $s<2$. For $s>2$ the jet geometry will change from parabolic to conical.

From the observational perspective several studies have concluded that the acceleration zone is located upstream from the radio core of the jet (thought to be a standing shock and the location at which the jet reaches its maximum Lorentz factor, e.g., \citealp{Marscher1995}) approximately at $10^5R_s$ from the BH, where $R_s$ is the Schwarzschild radius \citep{Marscher2008,Marscher2010}. Recent results on M87 suggest that the jet has a parabolic profile and accelerated up to the Bondi radius (which marks the sphere of gravitational influence of the BH, $\sim5\times10^5R_s$ also the location of HST-1), and then transitions to a conical geometry \citep{Asada2012,Nakamura2013,Asada2014}. Similar results for the acceleration profile of M87 have been obtained by wavelet analysis in \cite{Mertens2016}. This transition is thought to be caused by different profiles of external pressure making HST-1 a potential recollimation shock \citep{Stawarz2006,Levinson2017}. Results for Cygnus A suggest similar characteristics in jet structure and acceleration profile. The jet of Cygnus A is consistent with being externally confined and magnetically driven with the acceleration region extending up to $10^4R_s$ \citep{Boccardi2016}.

In this work, motivated by these recent results, we present a simple yet comprehensive toy model for the acceleration of blazar jets. Our goal is to create a simple framework on which both theorists and observers can build on in order to address more complex aspects of astrophysical jets. In Section \ref{toy_model} we present the toy model. In Section \ref{app} we apply our model to $\Gamma$ estimates of blazars and in Section \ref{disc_conc} we discuss the findings and conclusions of this work. { In the Appendix we discuss the possible application of the model to Gamma-Ray Bursts (GRBs).}
 
\section{Toy model}\label{toy_model}
\begin{figure}
 \centering
 \resizebox{\hsize}{!}{\includegraphics[scale=0.5]{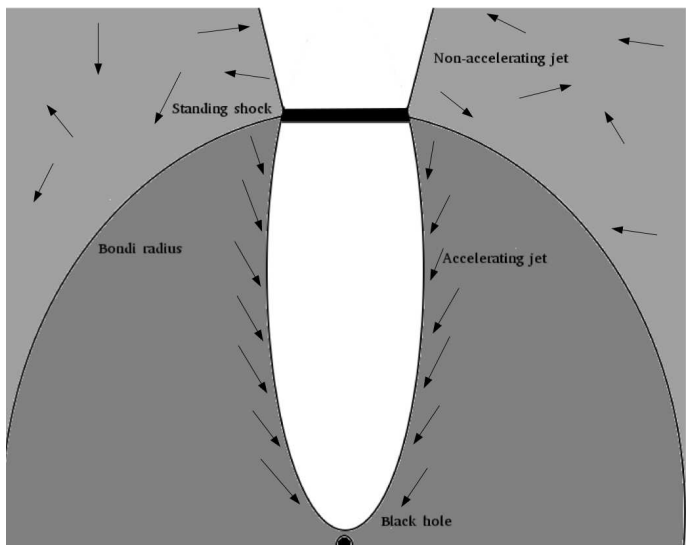} }
 \caption{Schematic of the toy model for the acceleration of astrophysical jets. The black arrows show the movement of the gas around the BH and jet.}
 \label{fig:toy}
\end{figure}
Considering the points raised above the toy model we propose is as follows. The jets are initially magnetically dominated and confined by external pressure having an initial parabolic geometry while accelerated over a large distance from the BH. The jet is accelerated through conversion of magnetic to kinetic energy until the two reach equipartition. The gas within the Bondi radius is forced to move inwards due to the gravitational pull of the BH. As expected from spherical accretion, the density and temperature of the gas will increase towards the BH creating a power-law profile for the density, and hence the power-law profile of the external pressure necessary to confine the jet (Bondi accretion has been found to be consistent with the observed luminosity of M87, \citealp{DiMatteo2003}). Outside the Bondi radius the gas is free to move in any direction, and thus the external pressure loses its profile and can no longer collimate the jet into a parabolic shape necessary for the acceleration. At the Bondi radius observations would suggest the existence of a recollimation shock \citep{Asada2012,Asada2014}, which in blazars would be the observed radio core of the jet \citep{Daly1988,Marscher2008-II}. { The formation of the shock could be due to the difference in the pressure profile of the surrounding medium \citep{Gomez1997,Barniol-Duran2017}. Such a shock is also expected to form if the external pressure gradient is $s<2$ \citep{Komissarov1997}.} The shock is the location where the jet reaches its maximum Lorentz factor since: (1) after the shock the jet is no longer collimated in a parabolic geometry and cannot be efficiently accelerated; and (2) the standing shock will inevitably decelerate the flow. Beyond the Bondi radius we have adopted a conical geometry as suggested by observations {(\citealp{Asada2012,Asada2014}, see Section \ref{disc_conc})}. The overall characteristics of the toy model are summarized in Fig.\ref{fig:toy}. In the equilibrium regime (where the jet is efficiently accelerated) the Lorentz factor grows as
\begin{equation}
\Gamma\approx\left(\frac{z}{\omega_{LC}}   \right)^{s/4},
\label{eq:gamma_max_orig}
\end{equation}
where $z$ is the distance from the BH, $s$ is the power-law index of the external pressure ($p\propto z^{-s}$), and $\omega_{LC}=c/\Omega=c/0.5\Omega_h$ is the cylindrical radius of the light cylinder, where $c$ is the speed of light, and $\Omega_h$ the angular velocity of the BH (\citealp{Komissarov2009,Lyubarsky2009,Boettcher2012-book} and references therein). According to the toy model the maximum $\Gamma$ is reached at the Bondi radius, i.e., $z=r_{\rm Bondi}=2GM/v_\infty^2$, where $G$ is the gravitational constant, $M$ the mass of the BH, and $v_\infty$ the sound speed at the Bondi radius. Then,
\begin{equation}
\Gamma_{\rm max}=\left(\frac{2GM 0.5 \Omega_h}{c v_\infty^2}   \right)^{s/4}.
\label{eq:gamma_max}
\end{equation}
The angular velocity of the BH is defined as
\begin{equation}
\Omega_h=f_{\Omega_h}(a)\frac{c^3}{2GM},
\label{eq:omega_h}
\end{equation}
where $f_{\Omega_h}(a)=a/(1+\sqrt{1-a^2})$, and $a$ is the dimensionless spin of the BH. Assuming a mean molecular weight $\mu=0.6$ and temperature $T=6.5\times10^6~K$ (consistent with observations of M87, \citealp{Narayan2011,Russell2015}) the sound speed at the Bondi radius becomes
\begin{equation}
v_\infty=10^{-3}c.
\label{eq:sound_c}
\end{equation}
Combining Eq. \ref{eq:gamma_max}, \ref{eq:omega_h}, and \ref{eq:sound_c},
\begin{equation}
\Gamma_{\rm max}=\left[5\times10^5f_{\Omega_h}(a) \right]^{s/4}.
\label{eq:gamma_max2}
\end{equation}
Equation \ref{eq:gamma_max2} is independent of the BH mass which is a necessary condition since similar mass BHs in different systems (i.e microquasars and GRBs) produce jets with up to two orders of magnitude different $\Gamma$. $\Gamma_{\rm max}$ depends only on the spin of the BH and the gradient of the external pressure, with $\Gamma$ having a stronger dependence on the latter. For example, assuming $s=a=0.9$, Eq. \ref{eq:gamma_max2} yields $\Gamma_{\rm max}\approx 17.24$. For a 10\% change in $a$ there is a $7.6\%$ change in $\Gamma_{\rm max}$, while a 10\% change in $s$ results in a 33\% change in $\Gamma_{\rm max}$. { Thus there is only a mild dependence of $\Gamma_{\rm max}$ on the spin.} 

\section{Application to blazar jets}\label{app}

\begin{figure}
 \centering
 \resizebox{\hsize}{!}{\includegraphics[scale=1]{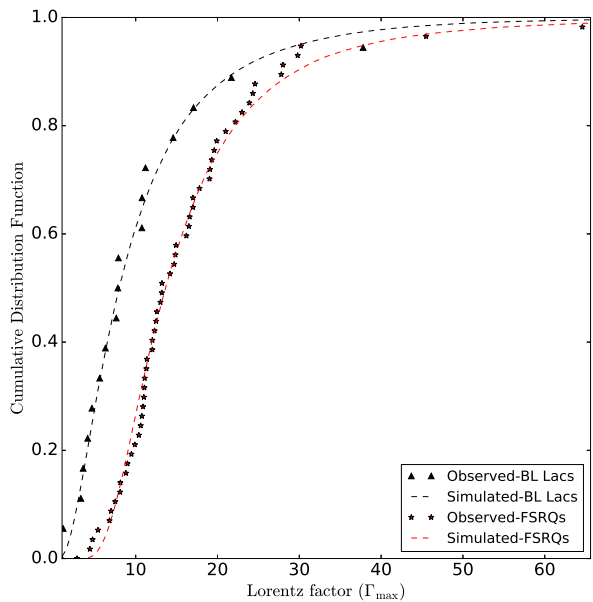} }
 \caption{Cumulative distribution function for the observed and simulated $\Gamma_{\rm max}$. Black triangles are for the observed BL Lacs and red stars for the observed FSRQs. The dashed black and red lines are the simulated sample for BL Lacs and FSRQs respectively.}
 \label{fig:pop_spin}
\end{figure}

\begin{table}
\setlength{\tabcolsep}{8pt}
\centering
  \caption{Parameters of the best-fit distributions of $a$, $s$ for BL Lacs and FSRQs. Columns: (1) Class, (2) parameter, (3) mean,  (4) standard deviation, (5) reduced $\chi^2$ value, (6) p-value of the reduced $\chi^2$, (7) p-value of the K-S test.}
 \label{table:distributions}  
\begin{tabular}{@{}lcccccr@{}}
 \hline
Class & Param. & $\mu$ & $\sigma$ & $\chi^2$ &$\chi^2$ & K-S \\
 &  &  &  &  & (\%) & (\%) \\ 
   \hline 
   \\
\hline
BL Lacs & & & &0.04 & 97.8& 95.6\\
\hline
 & $a$ & 0.937& 0.074& & & \\
& $s$ & 0.65 & 0.25& & & \\
\hline
FSRQs & & & &0.04 & 98.0& 96.2\\
\hline
 & $a$ &0.742 &0.163 & & & \\
& $s$ & 0.885& 0.175 & & & \\
\end{tabular}
\begin{threeparttable}
\begin{tablenotes}
\item[a] In both cases $a$ follows a beta distribution. $s$ follows a normal distribution for the BL Lacs and a log-normal for FSRQs. For the beta distribution the $\mu$ and $\sigma$ are defined as $\mu=\alpha/(\alpha+\beta)$, $\sigma^2=\alpha\beta/[(\alpha+\beta)^2(\alpha+\beta+1)]$ where $\alpha$, $\beta$ are the shape parameters. For the log-normal distribution the $\mu$ and $\sigma$ are defined as $\mu=\exp(loc+sc^2/2)$, $\sigma^2=(\exp(sc^2)-1)\exp(2l+sc^2)$ where $l$,$sc$ are the location and scale parameters respectively.
\end{tablenotes}
\end{threeparttable}

\end{table}

Studies on the spin and external pressure gradient of beamed sources are extremely rare and in the majority of cases unfeasible. The only source will available estimates for all three parameters that enter Eq. \ref{eq:gamma_max2} is M87. Studies of M87 have determined that the gradient of the external pressure has a power-law index of $s=0.6$  
\citep{Stawarz2006}; the maximum velocity of the jet at HST-1 is $\Gamma_{\rm max}=7.21\pm1.12$ \citep{Wang2009}; and the BH has a spin of $a\approx 0.98^{+0.012}_{-0.02}$ \citep{Feng2017}. Using any pair of the above parameters Eq. \ref{eq:gamma_max2} would yield the third within the uncertainties. Thus the model can {produce values consistent with} all three observed properties of the jet of M87.

Although we lack estimates of the $a$ and $s$ for blazars, we were able to use their observed $\Gamma$ (under the assumption that it is equal to $\Gamma_{\rm max}$) to constrain the distributions of the spin and the gradient of the external pressure. There are 75 blazar jets with available $\Gamma$ estimates (\citealp{Hovatta2009,Liodakis2017}, hereafter H09 and L17 respectively) 18 of which are BL Lac objects (BL Lacs) and 57 are flat spectrum radio quasars (FSRQ). These estimates are derived using variability Doppler factors (H09, L17) and apparent velocity estimates \citep{Lister2009-2,Lister2013}.
 
We assumed a distribution for $a$ and $s$ and use the observed $\Gamma_{\rm max}$ of blazars to constrain the optimal parameters for these distributions using a chi-square ($\chi^2$) minimization procedure. For $a$, since it is bounded between [0,1] a beta distribution is the natural choice\footnote{{Although different black holes are generally expected to have different spins, given the mild dependence of $\Gamma_{\rm max}$ on $a$ we also tested a delta function for the spin. The best-fit $a$ for both populations is $a\approx0.72$. Even with the fewer degrees of freedom, the beta distribution still yielded, albeit marginally, a better model according to the reduced $\chi^2$. The K-S test also favors the beta distribution over the delta function.}}. For $s$ we tested a normal, a log-normal and a uniform distribution. The distributions (and their parameters) that yielded the lowest reduced $\chi^2$ value are shown in Table \ref{table:distributions}. The results of the minimization were verified using the Kolmogorov-Smirnov (K-S) test\footnote{The K-S test yields the probability of two samples being drawn from the same distribution. We do not reject the null hypothesis for any p-value $>5\%$.}. Figure \ref{fig:pop_spin} shows the cumulative distribution function for the observed and simulated $\Gamma_{\rm max}$ for both blazar populations.

The best-fit distributions for $a$ and $s$ are different for BL Lacs and FSRQs. For the spin, BL Lacs have generally larger spins with a mean of $\mu=0.937$ while FSRQs have a mean of  $\mu=0.742$. It has been shown analytically that for spin values $a<0.6$ the BZ mechanism is no longer efficient \citep{Maraschi2012}.  In order to account for the observed $\gamma$-ray emission of blazars \cite{Maraschi2012} constrained the spin of blazars to $a>0.5$, possibly as high as $a\sim0.8$. Cosmological simulations of both BL Lacs and FSRQs have also determined that a sharp cut-off in the spin distribution of blazars is necessary in order to reproduce the number of observed sources in the Universe \citep{Gardner2014,Gardner2017}. Roughly 99.6\% of BL Lacs and 80.5\% of FSRQs in our sample have $a>0.6$. In addition, BL Lacs peak at $a\sim 0.9$ and FSRQs at $a\sim 0.8$  showing that our model naturally reproduces the results from different energetic and cosmological perspectives.

For the gradient of the external pressure, the BL Lacs follow a normal, while the FSRQs a log-normal distribution. The FSRQ distribution is also centered at, and extends to, higher values. This would suggest that, on average, the environment in the vicinity of the BHs of FSRQs is denser and therefore more gas-rich than the environment in the vicinity of the BHs in BL Lacs. Environmental conditions have been invoked in the past to explain the dichotomy between FR~I and FR~II type galaxies (the parent population of BL Lacs and FSRQs respectively). The results of the model would be consistent with evolutionary scenarios that attribute the differences in the two populations (BL Lacs \& FSRQs) to differences in their respective accretion rates (e.g., \citealp{Bottcher2002,Cavaliere2002,Ajello2014}). If this is the case, then the fact that BL Lacs show on average higher spins would suggest that their BHs were spun up in the past either by accretion of gas that is now depleted (suggesting that BL Lacs are more evolved blazars than FSRQs) or by gas-poor mergers (e.g., \citealp{Volonteri2005,Volonteri2007,Fanidakis2011}) suggesting a different evolutionary track than FSRQs. { The derived values for $s$ are swallower than predicted for Bondi accretion. They are, however, consistent with observations of M87 \citep{Stawarz2006} and are, for example, expected in the case were a ion torus supporting the jet is extending outwards from the supermassive black hole \citep{Rees1982}. }

In order to constrain the values for $a$ and $s$ for individual sources, we draw random values from the optimized distributions for $a$, $s$ for each population and minimize the square of the difference between observed $\Gamma_{\rm max}$ and  the expectation from the toy model ($[\Gamma_{\rm Obs}-\Gamma_{\rm model}]^2$). Table  \ref{table:estimates} in Appendix \ref{appendix} lists the optimal pairs of $a$,$s$ that reproduced the observed $\Gamma_{\rm max}$ after $10^5$ random draws. { It should be noted that given the mild dependence of $\Gamma_{\rm max}$ on the spin, the values of $s$ are better constraint. Nearby sources with large viewing angles (e.g., J1221+2813, L17) could be used to test the predictions of the model for the gradient of the external pressure.}

\section{Discussion and conclusions}\label{disc_conc}

For the application of our model on blazars we used $\Gamma$ estimates from radio observations. These estimates were derived using a wide range of observing frequencies from 2.6 to 43~GHz (H09, L17). The radio frequency necessary to probe the region where the $\Gamma_{\rm max}$ is achieved dependents on the properties of source. Results from the MOJAVE survey would suggest that more than half of the blazar jets show accelerating features at 15~GHz \citep{Homan2015,Lister2016}. Multiwavelength radio observations are then necessary to determine where and whether the $\Gamma_{\rm max}$ has been reached. For sources whose radio components show significant acceleration at the radio frequency where the Doppler factor (and hence $\Gamma$) was derived, the results of the model should be treated as lower limits.

{Beyond the Bondi radius we have assumed that the jet  has a conical geometry. This might not always be the case. Observations of M87 do support that scenario \citep{Asada2012,Asada2014}. MHD simulations have also shown that beyond the recollimation shock at the Bondi radius the jet could become conical, however, depending on the pressure and density profile of the medium outside the Bondi radius different geometries are possible. The resulting geometry could have an impact on the velocity profile of the jet at large scales (e.g., \citealp{Barniol-Duran2017}). Observations of additional AGN jet environments could give more insights on the fate of the jet beyond the Bondi radius.}

Throughout this work, we have assumed that the earliest the jet would reach $\sigma\sim 1$ is at the Bondi radius. It is, however, possible for the jet to cease being Poynting dominated before reaching the radio core. In such a case the results of the model should be treated as upper limits. We have also assumed that the jet comprises of one bulk flow. It is possible that the Lorentz factor can also change transversely along the jet. Observations of high synchrotron peaked (HSP) sources in the TeV band have shown variability timescales which would require much larger Doppler factors than the ones derived from radio observations. In order to explain this discrepancy, \cite{Ghisellini2005} suggested a spine-sheath configuration: a fast inner spine responsible for the high-energy emission and a slower outer sheath. In this configuration, \cite{Ghisellini2005} found that the spectral energy distribution of four sources can be well described if the sheath has a $\Gamma=[3,3.5]$ and the spine a $\Gamma=[15,17]$. If such is the case for blazar jets then the radio observations (which probe the scales at which the $\Gamma_{\rm max}$ is reached) will be dominated by emission from the sheath \citep{Sikora2016}. Although there are alternate hypothesis to the spine-sheath configuration that can fully explain the observed high energy emission without the need to invoke a faster bulk flow than the one derived from radio observations (e.g., magnetic reconnection, \citealp{Giannios2009,Giannios2010}) the model can be easily extended to incorporate different flow configurations.

In this work we have presented a simple toy model for the structure and acceleration of jets from supermassive black holes and its application using observed $\Gamma$ estimates of blazars. Our findings can be summarized as follows:
\begin{itemize}

\item Application to M87 showed that the model can produce { consistent values with all three properties} of the jet derived independently from observations.

\item BL Lacs have on average higher spins than FSRQs, with both populations having the vast majority of sources with $a>0.6$ consistent with energetic considerations for the efficiency of the BZ mechanism as well as cosmological simulations.

\item The results for the distribution of $s$ in BL Lacs and FSRQs would suggest that the BHs of the latter are, on average, in gas-richer and denser environments than the BHs of the former consistent with evolutionary models that attribute the differences of the two populations in their respective accretion rates.

\end{itemize}

Although there are many different aspects of the jets that have not been taken into account (MHD instabilities, energy conversion and dissipation mechanisms etc.), the fact that the model { can produce consistent results with the} observed properties of M87 as well as the general properties of blazars would suggest that it is a good first approximation on which a more complex and realistic model could be built.

\section*{Acknowledgments}
The author would like to thank the anonymous referee, Rodolfo Barniol Duran, Roger Blandford, Vasiliki Pavlidou, and Roger Romani for comments and discussions that helped improve this work.

\bibliographystyle{aa}
\bibliography{bibliography} 

\begin{thebibliography}{61}
\expandafter\ifx\csname natexlab\endcsname\relax\def\natexlab#1{#1}\fi

\bibitem[{{Ajello} {et~al.}(2014){Ajello}, {Romani}, {Gasparrini}, {Shaw},
  {Bolmer}, {Cotter}, {Finke}, {Greiner}, {Healey}, {King}, {Max-Moerbeck},
  {Michelson}, {Potter}, {Rau}, {Readhead}, {Richards}, \&
  {Schady}}]{Ajello2014}
{Ajello}, M., {Romani}, R.~W., {Gasparrini}, D., {et~al.} 2014, \apj, 780, 73

\bibitem[{{Asada} \& {Nakamura}(2012)}]{Asada2012}
{Asada}, K. \& {Nakamura}, M. 2012, \apjl, 745, L28

\bibitem[{{Asada} {et~al.}(2014){Asada}, {Nakamura}, {Doi}, {Nagai}, \&
  {Inoue}}]{Asada2014}
{Asada}, K., {Nakamura}, M., {Doi}, A., {Nagai}, H., \& {Inoue}, M. 2014,
  \apjl, 781, L2

\bibitem[{{Barniol Duran} {et~al.}(2017){Barniol Duran}, {Tchekhovskoy}, \&
  {Giannios}}]{Barniol-Duran2017}
{Barniol Duran}, R., {Tchekhovskoy}, A., \& {Giannios}, D. 2017, \mnras, 469,
  4957

\bibitem[{{Begelman}(2014)}]{Begelman2014}
{Begelman}, M.~C. 2014, ArXiv e-prints

\bibitem[{{Belczynski} {et~al.}(2010){Belczynski}, {Bulik}, {Fryer}, {Ruiter},
  {Valsecchi}, {Vink}, \& {Hurley}}]{Belczynski2010}
{Belczynski}, K., {Bulik}, T., {Fryer}, C.~L., {et~al.} 2010, \apj, 714, 1217

\bibitem[{{Blandford} \& {Znajek}(1977)}]{Blandford1977}
{Blandford}, R.~D. \& {Znajek}, R.~L. 1977, \mnras, 179, 433

\bibitem[{{Boccardi} {et~al.}(2016){Boccardi}, {Krichbaum}, {Bach}, {Mertens},
  {Ros}, {Alef}, \& {Zensus}}]{Boccardi2016}
{Boccardi}, B., {Krichbaum}, T.~P., {Bach}, U., {et~al.} 2016, \aap, 585, A33

\bibitem[{{Boettcher} {et~al.}(2012){Boettcher}, {Harris}, \&
  {Krawczynski}}]{Boettcher2012-book}
{Boettcher}, M., {Harris}, D.~E., \& {Krawczynski}, H. 2012, {Relativistic Jets
  from Active Galactic Nuclei}

\bibitem[{{B{\"o}ttcher} \& {Dermer}(2002)}]{Bottcher2002}
{B{\"o}ttcher}, M. \& {Dermer}, C.~D. 2002, \apj, 564, 86

\bibitem[{{Cavaliere} \& {D'Elia}(2002)}]{Cavaliere2002}
{Cavaliere}, A. \& {D'Elia}, V. 2002, \apj, 571, 226

\bibitem[{{Curtis}(1918)}]{Curtis1918}
{Curtis}, H.~D. 1918, Publications of Lick Observatory, 13, 9

\bibitem[{{Daly} \& {Marscher}(1988)}]{Daly1988}
{Daly}, R.~A. \& {Marscher}, A.~P. 1988, \apj, 334, 539

\bibitem[{{Di Matteo} {et~al.}(2003){Di Matteo}, {Allen}, {Fabian}, {Wilson},
  \& {Young}}]{DiMatteo2003}
{Di Matteo}, T., {Allen}, S.~W., {Fabian}, A.~C., {Wilson}, A.~S., \& {Young},
  A.~J. 2003, \apj, 582, 133

\bibitem[{{Fanidakis} {et~al.}(2011){Fanidakis}, {Baugh}, {Benson}, {Bower},
  {Cole}, {Done}, \& {Frenk}}]{Fanidakis2011}
{Fanidakis}, N., {Baugh}, C.~M., {Benson}, A.~J., {et~al.} 2011, \mnras, 410,
  53

\bibitem[{{Feng} \& {Wu}(2017)}]{Feng2017}
{Feng}, J. \& {Wu}, Q. 2017, \mnras, 470, 612

\bibitem[{{Gardner} \& {Done}(2014)}]{Gardner2014}
{Gardner}, E. \& {Done}, C. 2014, \mnras, 438, 779

\bibitem[{{Gardner} \& {Done}(2017)}]{Gardner2017}
{Gardner}, E. \& {Done}, C. 2017, ArXiv e-prints

\bibitem[{{Ghisellini} {et~al.}(2005){Ghisellini}, {Tavecchio}, \&
  {Chiaberge}}]{Ghisellini2005}
{Ghisellini}, G., {Tavecchio}, F., \& {Chiaberge}, M. 2005, \aap, 432, 401

\bibitem[{{Giannios} {et~al.}(2009){Giannios}, {Uzdensky}, \&
  {Begelman}}]{Giannios2009}
{Giannios}, D., {Uzdensky}, D.~A., \& {Begelman}, M.~C. 2009, \mnras, 395, L29

\bibitem[{{Giannios} {et~al.}(2010){Giannios}, {Uzdensky}, \&
  {Begelman}}]{Giannios2010}
{Giannios}, D., {Uzdensky}, D.~A., \& {Begelman}, M.~C. 2010, \mnras, 402, 1649

\bibitem[{{G{\'o}mez} {et~al.}(1997){G{\'o}mez}, {Mart{\'{\i}}}, {Marscher},
  {Ib{\'a}{\~n}ez}, \& {Alberdi}}]{Gomez1997}
{G{\'o}mez}, J.~L., {Mart{\'{\i}}}, J.~M., {Marscher}, A.~P., {Ib{\'a}{\~n}ez},
  J.~M., \& {Alberdi}, A. 1997, \apjl, 482, L33

\bibitem[{{Homan} {et~al.}(2015){Homan}, {Lister}, {Kovalev}, {Pushkarev},
  {Savolainen}, {Kellermann}, {Richards}, \& {Ros}}]{Homan2015}
{Homan}, D.~C., {Lister}, M.~L., {Kovalev}, Y.~Y., {et~al.} 2015, \apj, 798,
  134

\bibitem[{{Hovatta} {et~al.}(2009){Hovatta}, {Valtaoja}, {Tornikoski}, \&
  {L{\"a}hteenm{\"a}ki}}]{Hovatta2009}
{Hovatta}, T., {Valtaoja}, E., {Tornikoski}, M., \& {L{\"a}hteenm{\"a}ki}, A.
  2009, \aap, 494, 527

\bibitem[{{Komissarov} {et~al.}(2007){Komissarov}, {Barkov}, {Vlahakis}, \&
  {K{\"o}nigl}}]{Komissarov2007}
{Komissarov}, S.~S., {Barkov}, M.~V., {Vlahakis}, N., \& {K{\"o}nigl}, A. 2007,
  \mnras, 380, 51

\bibitem[{{Komissarov} \& {Falle}(1997)}]{Komissarov1997}
{Komissarov}, S.~S. \& {Falle}, S.~A.~E.~G. 1997, \mnras, 288, 833

\bibitem[{{Komissarov} {et~al.}(2010){Komissarov}, {Vlahakis}, \&
  {K{\"o}nigl}}]{Komissarov2010}
{Komissarov}, S.~S., {Vlahakis}, N., \& {K{\"o}nigl}, A. 2010, \mnras, 407, 17

\bibitem[{{Komissarov} {et~al.}(2009){Komissarov}, {Vlahakis}, {K{\"o}nigl}, \&
  {Barkov}}]{Komissarov2009}
{Komissarov}, S.~S., {Vlahakis}, N., {K{\"o}nigl}, A., \& {Barkov}, M.~V. 2009,
  \mnras, 394, 1182

\bibitem[{{Levinson} \& {Globus}(2017)}]{Levinson2017}
{Levinson}, A. \& {Globus}, N. 2017, \mnras, 465, 1608

\bibitem[{{Liodakis} {et~al.}(2017){Liodakis}, {Marchili}, {Angelakis},
  {Fuhrmann}, {Nestoras}, {Myserlis}, {Karamanavis}, {Krichbaum}, {Sievers},
  {Ungerechts}, \& {Zensus}}]{Liodakis2017}
{Liodakis}, I., {Marchili}, N., {Angelakis}, E., {et~al.} 2017, \mnras, 466,
  4625

\bibitem[{{Lister} {et~al.}(2013){Lister}, {Aller}, {Aller}, {Homan},
  {Kellermann}, {Kovalev}, {Pushkarev}, {Richards}, {Ros}, \&
  {Savolainen}}]{Lister2013}
{Lister}, M.~L., {Aller}, M.~F., {Aller}, H.~D., {et~al.} 2013, \aj, 146, 120

\bibitem[{{Lister} {et~al.}(2016){Lister}, {Aller}, {Aller}, {Homan},
  {Kellermann}, {Kovalev}, {Pushkarev}, {Richards}, {Ros}, \&
  {Savolainen}}]{Lister2016}
{Lister}, M.~L., {Aller}, M.~F., {Aller}, H.~D., {et~al.} 2016, ArXiv e-prints

\bibitem[{{Lister} {et~al.}(2009){Lister}, {Cohen}, {Homan}, {Kadler},
  {Kellermann}, {Kovalev}, {Ros}, {Savolainen}, \& {Zensus}}]{Lister2009-2}
{Lister}, M.~L., {Cohen}, M.~H., {Homan}, D.~C., {et~al.} 2009, \aj, 138, 1874

\bibitem[{{Lyubarsky}(2009)}]{Lyubarsky2009}
{Lyubarsky}, Y. 2009, \apj, 698, 1570

\bibitem[{{Lyubarsky}(2010)}]{Lyubarsky2010}
{Lyubarsky}, Y.~E. 2010, \mnras, 402, 353

\bibitem[{{Maraschi} {et~al.}(2012){Maraschi}, {Colpi}, {Ghisellini}, {Perego},
  \& {Tavecchio}}]{Maraschi2012}
{Maraschi}, L., {Colpi}, M., {Ghisellini}, G., {Perego}, A., \& {Tavecchio}, F.
  2012, in Journal of Physics Conference Series, Vol. 355, Journal of Physics
  Conference Series, 012016

\bibitem[{{Marscher}(1995)}]{Marscher1995}
{Marscher}, A.~P. 1995, Proceedings of the National Academy of Science, 92,
  11439

\bibitem[{{Marscher}(2008)}]{Marscher2008-II}
{Marscher}, A.~P. 2008, in Astronomical Society of the Pacific Conference
  Series, Vol. 386, Extragalactic Jets: Theory and Observation from Radio to
  Gamma Ray, ed. T.~A. {Rector} \& D.~S. {De Young}, 437

\bibitem[{{Marscher} {et~al.}(2008){Marscher}, {Jorstad}, {D'Arcangelo},
  {Smith}, {Williams}, {Larionov}, {Oh}, {Olmstead}, {Aller}, {Aller},
  {McHardy}, {L{\"a}hteenm{\"a}ki}, {Tornikoski}, {Valtaoja}, {Hagen-Thorn},
  {Kopatskaya}, {Gear}, {Tosti}, {Kurtanidze}, {Nikolashvili}, {Sigua},
  {Miller}, \& {Ryle}}]{Marscher2008}
{Marscher}, A.~P., {Jorstad}, S.~G., {D'Arcangelo}, F.~D., {et~al.} 2008, \nat,
  452, 966

\bibitem[{{Marscher} {et~al.}(2010){Marscher}, {Jorstad}, {Larionov}, {Aller},
  {Aller}, {L{\"a}hteenm{\"a}ki}, {Agudo}, {Smith}, {Gurwell}, {Hagen-Thorn},
  {Konstantinova}, {Larionova}, {Larionova}, {Melnichuk}, {Blinov},
  {Kopatskaya}, {Troitsky}, {Tornikoski}, {Hovatta}, {Schmidt}, {D'Arcangelo},
  {Bhattarai}, {Taylor}, {Olmstead}, {Manne-Nicholas}, {Roca-Sogorb},
  {G{\'o}mez}, {McHardy}, {Kurtanidze}, {Nikolashvili}, {Kimeridze}, \&
  {Sigua}}]{Marscher2010}
{Marscher}, A.~P., {Jorstad}, S.~G., {Larionov}, V.~M., {et~al.} 2010, \apjl,
  710, L126

\bibitem[{{Mertens} {et~al.}(2016){Mertens}, {Lobanov}, {Walker}, \&
  {Hardee}}]{Mertens2016}
{Mertens}, F., {Lobanov}, A.~P., {Walker}, R.~C., \& {Hardee}, P.~E. 2016,
  \aap, 595, A54

\bibitem[{{M{\'e}sz{\'a}ros} \& {Rees}(2001)}]{Menzaros2001}
{M{\'e}sz{\'a}ros}, P. \& {Rees}, M.~J. 2001, \apjl, 556, L37

\bibitem[{{Nakamura} \& {Asada}(2013)}]{Nakamura2013}
{Nakamura}, M. \& {Asada}, K. 2013, \apj, 775, 118

\bibitem[{{Narayan} \& {Fabian}(2011)}]{Narayan2011}
{Narayan}, R. \& {Fabian}, A.~C. 2011, \mnras, 415, 3721

\bibitem[{{Piran}(2004)}]{Piran2004}
{Piran}, T. 2004, Reviews of Modern Physics, 76, 1143

\bibitem[{{Rees} {et~al.}(1982){Rees}, {Begelman}, {Blandford}, \&
  {Phinney}}]{Rees1982}
{Rees}, M.~J., {Begelman}, M.~C., {Blandford}, R.~D., \& {Phinney}, E.~S. 1982,
  \nat, 295, 17

\bibitem[{{Russell} {et~al.}(2015){Russell}, {Fabian}, {McNamara}, \&
  {Broderick}}]{Russell2015}
{Russell}, H.~R., {Fabian}, A.~C., {McNamara}, B.~R., \& {Broderick}, A.~E.
  2015, \mnras, 451, 588

\bibitem[{{Sapountzis} \& {Vlahakis}(2013)}]{Sapountzis2013}
{Sapountzis}, K. \& {Vlahakis}, N. 2013, \mnras, 434, 1779

\bibitem[{{Schaerer} \& {Maeder}(1992)}]{Schaerer1992}
{Schaerer}, D. \& {Maeder}, A. 1992, \aap, 263, 129

\bibitem[{{Sikora} {et~al.}(2016){Sikora}, {Rutkowski}, \&
  {Begelman}}]{Sikora2016}
{Sikora}, M., {Rutkowski}, M., \& {Begelman}, M.~C. 2016, \mnras, 457, 1352

\bibitem[{{Stawarz} {et~al.}(2006){Stawarz}, {Aharonian}, {Kataoka},
  {Ostrowski}, {Siemiginowska}, \& {Sikora}}]{Stawarz2006}
{Stawarz}, {\L}., {Aharonian}, F., {Kataoka}, J., {et~al.} 2006, \mnras, 370,
  981

\bibitem[{{Tchekhovskoy} {et~al.}(2008){Tchekhovskoy}, {McKinney}, \&
  {Narayan}}]{Tchekhovskoy2008}
{Tchekhovskoy}, A., {McKinney}, J.~C., \& {Narayan}, R. 2008, \mnras, 388, 551

\bibitem[{{Tchekhovskoy} {et~al.}(2010){Tchekhovskoy}, {Narayan}, \&
  {McKinney}}]{Tchekhovskoy2010-II}
{Tchekhovskoy}, A., {Narayan}, R., \& {McKinney}, J.~C. 2010, \na, 15, 749

\bibitem[{{Vlahakis}(2004)}]{Vlahakis2004-II}
{Vlahakis}, N. 2004, \apss, 293, 67

\bibitem[{{Vlahakis}(2015)}]{Vlahakis2015-book}
{Vlahakis}, N. 2015, in Astrophysics and Space Science Library, Vol. 414, The
  Formation and Disruption of Black Hole Jets, ed. I.~{Contopoulos},
  D.~{Gabuzda}, \& N.~{Kylafis}, 177

\bibitem[{{Vlahakis} \& {K{\"o}nigl}(2003)}]{Vlahakis2003}
{Vlahakis}, N. \& {K{\"o}nigl}, A. 2003, \apj, 596, 1104

\bibitem[{{Vlahakis} \& {K{\"o}nigl}(2004)}]{Vlahakis2004}
{Vlahakis}, N. \& {K{\"o}nigl}, A. 2004, \apj, 605, 656

\bibitem[{{Volonteri} {et~al.}(2005){Volonteri}, {Madau}, {Quataert}, \&
  {Rees}}]{Volonteri2005}
{Volonteri}, M., {Madau}, P., {Quataert}, E., \& {Rees}, M.~J. 2005, \apj, 620,
  69

\bibitem[{{Volonteri} {et~al.}(2007){Volonteri}, {Sikora}, \&
  {Lasota}}]{Volonteri2007}
{Volonteri}, M., {Sikora}, M., \& {Lasota}, J.-P. 2007, \apj, 667, 704

\bibitem[{{Wang} \& {Zhou}(2009)}]{Wang2009}
{Wang}, C.-C. \& {Zhou}, H.-Y. 2009, \mnras, 395, 301

\bibitem[{{Woosley}(1993)}]{Woosley1993}
{Woosley}, S.~E. 1993, \apj, 405, 273

\end{thebibliography}

\appendix

\section{Application to gamma-ray bursts}

Although the model has been constructed based on observations of jets from supermassive black holes, it would be interesting to explore its application to GRBs. Within the collapsar model, \citep{Woosley1993} the stellar envelope could assume the role of the confining external medium collimating the flow. The jet would then be accelerated in a parabolic geometry until it breaks free from the star envelope to the ISM \citep{Menzaros2001}. According to \cite{Menzaros2001} $\Gamma$ would grow as $\Gamma\propto z^{1/2}$. If this is the case, our model can easily produce $\Gamma_{\rm max}$ of a few hundreds consistent with the $\Gamma$ seen in GRBs (e.g., \citealp{Piran2004,Begelman2014}). However, the toy model assumes that the end of acceleration takes place at the Bondi radius. In the case of GRBs the end of the jet acceleration would be at the radius ($R_\star$) of the progenitor star. Then $z$ in Eq. \ref{eq:gamma_max_orig} should be substituted with $R_\star$, $\Gamma=\left(0.5R_\star f_{\Omega_h}(a)c^2/2GM\right)^{s/4}=\left(0.25R_\star f_{\Omega_h}(a)c^2k/GM_\star\right)^{s/4}$, where $k$ is the mass ratio of the progenitor star to the resulting BH and depends on the metallicity of the progenitor \citep{Belczynski2010}. Using the mass-to-radius relation for Wolf-rayet stars, $R_\star/R_\odot=10^{-n}(M_\star/M_\odot)^m$, where $n=0.6629$ and $m=0.5840$ \citep{Schaerer1992} $\Gamma_{\rm max}$ becomes, 
\begin{equation}
\Gamma_{\rm max}=\left(\frac{0.25c^2k10^{-n/m}}{G}R_\star^{1-1/m} f_{\Omega_h}(a)\right)^{s/4}
\end{equation}
For similar metallicity progenitors $\Gamma_{\rm max}$ would depend on the spin of the resulting BH, the pressure profile within the stellar envelope, and the radius of the progenitor star, $\Gamma_{\rm max}\propto\left(R_\star^{1-1/m} f_{\Omega_h}(a)\right)^{s/4}$. RMHD simulations of GRBs have shown $\Gamma_{\rm max}$ to have the same dependences \citep{Tchekhovskoy2008}. However, it is also possible for the jet to experience rarefaction acceleration \citep{Tchekhovskoy2010-II,Komissarov2010,Sapountzis2013} when exiting the envelope of the progenitor star. If this is the case, the model could only be used to set the initial conditions of the rarefaction acceleration of the jet outside the progenitor star. The fact that the prediction of the model is in agreement with simulations shows some promise, although further investigation into whether this or a similar model is indeed applicable to GRBs is doubtless necessary.

\section{Spin and external pressure gradient estimates}\label{appendix}

\onecolumn
\begin{longtable}{lcccccrr}
\caption{\label{table:estimates} Spin and external pressure gradient estimates for blazars. Columns: (1) Name as given in H09, L17, (2) Alternative name, (3) Class (B is for BL Lacs, F for FSRQs), (4) Lorentz factor, (5) Spin, (6) External pressure gradient, (7) Reference for the Lorentz factor estimate.}\\
\hline\hline
Name & Alt-name &Class &$\Gamma$ & $a$ & $s$ & Ref. \\
\hline
\endfirsthead
\caption{continued.}\\
\hline\hline
Name & Alt-name &Class &$\Gamma$ & $a$ & $s$ & Ref. \\
\hline
\endhead
\hline
\endfoot
J0003-066 & NRAO  5 & B & 3.3 & 0.9 & 0.38 & H09  \\
J0016+731 & - & F & 6.8 & 0.59 & 0.64 & H09  \\
J0102+5824 & 0059+5808 & F & 12.0 & 0.6 & 0.83 & L17 & \\ 
J0106+013 & OC 012 & F & 27.8 & 0.82 & 1.07 & H09  \\
J0136+4751 & 0133+476 & F & 9.5 & 0.49 & 0.76 & L17  \\
J0202+149 & 4C 15.05 & F & 9.9 & 0.61 & 0.76 & H09  \\
J0212+735 & - & F & 7.5 & 0.57 & 0.67 & H09  \\
J0217+0144 & PKS 0215+015 & F & 19.1 & 0.75 & 0.96 & L17  \\
J0224+671 & - & F & 12.5 & 0.6 & 0.84 & H09  \\
J0237+2848 & 4C 28.07 & F & 14.9 & 0.75 & 0.88 & L17  \\
J0238+1636 & 0235+164 & B & 14.6 & 0.96 & 0.83 & L17  \\
J0333+321 & NRAO 140 & F & 14.7 & 0.9 & 0.85 & H09  \\
J0336-019 & CTA 026 & F & 23.0 & 0.78 & 1.01 & H09  \\
J0359+5057 & 0355+50 & F & 13.2 & 0.75 & 0.84 & L17  \\
J0423-0120 & PKS 0420-014 & F & 22.2 & 0.48 & 1.05 & L17  \\
J0458-020 & PKS 0458-020 & F & 16.2 & 0.89 & 0.88 & H09  \\
J0530+1331 & PKS 0528+134 & F & 10.8 & 0.86 & 0.76 & L17  \\
J0552+398 & DA 193 & F & 12.6 & 0.8 & 0.82 & H09  \\
J0605-085 & PKS 0605-085 & F & 30.2 & 0.54 & 1.15 & H09  \\
J0642+449 & OH 471 & F & 5.4 & 0.79 & 0.54 & H09  \\
J0721+7120 & PKS 0716+714 & B & 10.8 & 0.93 & 0.75 & L17  \\
J0736+017 & - & F & 17.0 & 0.8 & 0.91 & H09  \\
J0738+1742 & 0735+178 & B & 3.6 & 1.0 & 0.39 & L17  \\
J0754+100 & OI 090.4 & B & 21.7 & 0.9 & 0.97 & H09  \\
J0804+499 & - & F & 17.8 & 0.58 & 0.96 & H09  \\
J0808-0751 & 0805-077 & F & 24.3 & 0.94 & 1.0 & L17  \\
J0818+4222 & 0814+425 & B & 4.1 & 0.93 & 0.45 & L17  \\
J0827+243 & OJ 248 & F & 23.9 & 0.62 & 1.05 & H09  \\
J0836+710 & 4C 71.07 & F & 28.0 & 0.38 & 1.16 & H09  \\
J0854+2006 & OJ 287 & B & 7.6 & 0.82 & 0.65 & L17  \\
J0920+4441 & S4 0917+449 & F & 2.8 & 0.67 & 0.37 & L17  \\
J0923+392 & 4C 39.25 & F & 4.4 & 0.53 & 0.5 & H09  \\
J0945+408 & 4C 40.24 & F & 29.8 & 0.76 & 1.1 & H09  \\
J0958+6533 & 0954+658 & B & 7.9 & 0.98 & 0.64 & L17  \\
J1055+018 & OL 093 & F & 11.1 & 0.57 & 0.8 & H09  \\
J1104+3812 & PKS 1101+384 & B & 1.1 & 1.0 & 0.04 & L17  \\
J1130-1449 & 1127-145 & F & 13.0 & 0.26 & 0.93 & L17  \\
J1159+2914 & PKS 1156+295 & F & 16.6 & 0.65 & 0.93 & L17  \\
J1221+2813 & QSOB1219+285 & B & 4.6 & 0.99 & 0.47 & L17  \\
J1222+216 & PKS 1222+216 & F & 45.5 & 0.57 & 1.28 & H09  \\
J1229+0203 & 3C 273 & F & 12.0 & 0.69 & 0.82 & L17  \\
J1256-0547 & 3C 279 & F & 12.3 & 0.65 & 0.83 & L17  \\
J1310+3220 & 1308+326 & B & 17.1 & 0.99 & 0.87 & L17  \\
J1324+224 & - & F & 10.9 & 0.82 & 0.77 & H09  \\
J1332-0509 & PKS 1329-049 & F & 11.1 & 0.39 & 0.83 & L17  \\
J1413+135 & - & B & 6.3 & 0.98 & 0.57 & H09  \\
J1504+1029 & OR 103 & F & 11.4 & 0.7 & 0.8 & L17  \\
J1512-0905 & PKS 1510-089 & F & 19.0 & 0.81 & 0.95 & L17  \\
J1538+149 & 4C 14.60 & B & 11.2 & 1.0 & 0.74 & H09  \\
J1606+106 & 4C 10.45 & F & 19.6 & 0.79 & 0.96 & H09  \\
J1611+343 & DA 406 & F & 14.2 & 0.71 & 0.87 & H09  \\
J1635+3808 & 4C 38.41 & F & 14.9 & 0.82 & 0.87 & L17  \\
J1637+574 & OS 562 & F & 11.0 & 0.63 & 0.79 & H09  \\
J1642+3948 & 3C 345 & F & 11.3 & 0.53 & 0.82 & L17  \\
J1730-130 & NRAO 530 & F & 64.6 & 0.52 & 1.41 & H09  \\
J1751+0939 & PKS 1749+096 & B & 7.8 & 0.98 & 0.64 & L17  \\
J1800+7828 & S3 1803+784 & B & 10.8 & 0.82 & 0.76 & L17  \\
J1807+698 & 3C 371.0 & B & 1.0 & 0.92 & 0.0 & H09  \\
J1823+568 & 4C 56.27 & B & 37.8 & 0.91 & 1.15 & H09  \\
J1828+487 & 3C 380 & F & 19.3 & 0.82 & 0.95 & H09  \\
J1848+3219 & TXS 1846+322 & F & 7.0 & 0.39 & 0.68 & L17  \\
J1849+6705 & S4 1849+670 & F & 17.0 & 0.92 & 0.89 & L17  \\
J1928+738 & 4C 73.18 & F & 19.9 & 0.79 & 0.96 & H09  \\
J2005+403 & - & F & 21.0 & 0.88 & 0.97 & H09  \\
J2025-0735 & PKS 2022-077 & F & 24.6 & 0.91 & 1.01 & L17  \\
J2121+053 & - & F & 13.2 & 0.75 & 0.84 & H09  \\
J2134+004 & OX 057 & F & 9.0 & 0.67 & 0.72 & H09  \\
J2143+1743 & PKS 2141+175 & F & 4.7 & 0.88 & 0.49 & L17  \\
J2201+315 & 4C 31.63 & F & 8.1 & 0.4 & 0.72 & H09  \\
J2202+4216 & BL Lacertae & B & 5.6 & 0.97 & 0.53 & L17  \\
J2223-052 & 3C 446 & F & 16.5 & 0.9 & 0.89 & H09  \\
J2227-088 & - & F & 8.8 & 0.86 & 0.69 & H09  \\
J2229-0832 & 2227-088 & F & 10.6 & 0.74 & 0.77 & L17  \\
J2232+1143 & CTA 102 & F & 8.1 & 0.46 & 0.72 & L17  \\
J2253+1608 & 3C 454.3 & F & 10.4 & 0.77 & 0.76 & L17  \\
\end{longtable}

\label{lastpage}

\end{document}